\documentclass[twocolumn,prb,aps]{revtex4}

 \usepackage[dvips]{graphicx}
 \usepackage[dvips]{graphics}

\begin{document}
\title{Odd and even magnetic resonant modes in highly overdoped
$\bf Bi_2 Sr_2 Ca Cu_2 O_{8+\delta}$ }

\author{
L. Capogna$^{1,2,3,\ast}$, B. Fauqu\'e$^{4}$, Y. Sidis$^4$, 
C. Ulrich$^3$, P. Bourges$^4$, S. Pailh\`es$^5$, A. Ivanov$^2$,
J.L. Tallon$^6$, B. Liang$^{3,\ast\ast}$, C.T. Lin$^3$, A.I. Rykov $^7$, and B.
Keimer$^3$}

\affiliation{
$^1$ INFM-CNR, 6 Rue J. Horowitz, 38042 Grenoble cedex 9, France\\
$^2$ Institut Laue-Langevin, 6 Rue J. Horowitz, 38042 Grenoble cedex 9, France\\
$^3$ MPI f\"ur Festk\"orperforschung, Heisenbergstr. 1, 70569 Stuttgart, Germany\\
$^4$ Laboratoire L\'{e}on Brillouin, CEA-CNRS, CEA-Saclay, 91191 Gif sur Yvette, France\\
$^5$ LNS, ETH Z\"urich and Paul Scherrer Institute, 5232 Villigen PSI, Switzerland\\
$^6$ Industrial Research Limited and Victoria University, P.O.
31310, Lower Hutt, New Zealand\\
$^7$ Department of Applied Chemistry, University of Tokyo, Hongo 7-3-1, Bunkyo-ku, Tokyo 113-8656, Japan
}
% Graduate School of Engineering, University of Tokyo, Hongo 7-3-1, Bunkyo-ku, Tokyo 113-8656, Japan
%}

%  25.40.Fq Inelastic neutron scattering
%74.72.Hs    Bi-based cuprates
% 74.25.Ha Superconductivity Magnetic properties

\begin{abstract}
We present inelastic neutron scattering data on highly overdoped
$\rm Bi_2 Sr_2 Ca Cu_2 O_{8+\delta}$ single crystals with
superconducting transition temperature $T_C=70$ K and, for
comparison, a nearly optimally doped crystal with $T_C=87$ K. In
both samples, magnetic resonant modes with odd and even symmetry
under exchange of the two CuO$_2$ layers in the unit cell are
observed. In the overdoped sample, the linewidth of the odd mode
is reduced compared with the optimally doped sample. This
finding is discussed in conjunction with recent evidence for
intrinsic inhomogeneities in this compound. The data on odd and
even resonant excitations are otherwise fully consistent with
trends established on the basis of data on YBa$_2$Cu$_3$O$_{6+x}$.
This confirms the universality of these findings and extends them
into the highly overdoped regime of the phase diagram.
\end{abstract}

\maketitle

In high-$T_C$ superconducting (SC) copper oxides, the experimental
study of low-energy excitations is essential to building and
testing microscopic models incorporating strong electronic
correlations. Angle-resolved photoemission spectroscopy (ARPES)
and inelastic neutron scattering (INS) are complementary
momentum-resolved techniques to probe charge and spin excitation
spectra, respectively. Comparison of ARPES and INS data is
expected to shed light on the interaction between spin and charge
excitations, which according to many models is at the root of the
mechanism of high-$T_C$ superconductivity \cite{eschrig}. Among
the different cuprate families, $\rm Bi_2 Sr_2 Ca Cu_2
O_{8+\delta}$ (Bi2212) is particularly suitable to carry out such
a comparative study. Due to the high quality of its surface, this
system has been widely investigated by ARPES \cite{campuzano}.
However, INS measurements were prohibited for a long time because
of the small size of the single crystals.

This problem has recently been overcome thanks to the improvement
in neutron flux on triple axis spectrometers, and to the use of
arrays of co-aligned single crystals \cite{fong,he,ybco}. ARPES
studies have shown that in the underdoped and optimally doped
regimes of the Bi2212 phase diagram, the system displays non-Fermi
liquid properties with incoherent charge transport. The charge
excitation spectrum displays pronounced anomalies around the
$(\pi/a,0)$ and $(0, \pi /a)$ wave vectors in the SC state: the
so-called peak-dip-hump feature. At the same time, a spin-triplet
excitation referred to as the magnetic resonant mode has been
observed by INS in optimally doped \cite{fong} and slightly
overdoped \cite{he} Bi2212 at the planar wave vector $(\pi/a,
\pi/a)$ approximately connecting the wave vectors at which the
peak-dip-hump feature is observed in ARPES, and at a
characteristic energy $E_r$=43 meV that corresponds to the energy
difference between the peak and the dip. As its counterpart in
ARPES, this mode is only observed in the superconducting state.
These observations point to a strong scattering process involving
spin-charge coupling.

Extensive INS studies of the magnetic resonant mode have been
performed in $\rm YBa_2Cu_3O_{6+x}$ (Y123) \cite{ybco}, which shares
its bilayer structure with Bi2212. In particular, it was shown that
magnetic interactions between the two CuO$_2$ layers in a bilayer
unit lead to the formation of two non-degenerate modes characterized
by even and odd symmetries with respect to exchange of the layers
\cite{pailhes1,pailhes3}. The relative spectral weight of the two modes
allows incisive tests of microscopic models of the resonant mode as
well as a determination of the bulk superconducting energy gap as a
function of doping \cite{pailhes2}. However, several drawbacks of
the Y123 system make similar work on other families of
high-temperature superconductors highly desirable. First, owing to
problems with surface stability only limited ARPES data are
available on this system \cite{lu,borisenko}, so that a quantitative
comparison between ARPES and neutron scattering data has thus far
not been carried out. Further, the Y123 crystal structure contains
CuO chains, an electronically active but non-generic structural
element. Although magnetic excitations originating from the CuO
chains have thus far not been clearly identified, the influence of
the chains on magnetic excitations in the CuO$_2$ planes is a matter
of current debate \cite{manske,yamase}. Finally, even with Ca
substitution the doping levels accessible in Y123 are limited to the
slightly overdoped regime.

Here we report the observation of even and odd magnetic excitations
 in the superconducting state of optimally doped and heavily overdoped Bi2212. 
The data encompass a doping level of $\delta=0.21$ that has thus far only
been reached in neutron scattering experiments on
La$_{2-x}$Sr$_x$CuO$_4$ (Ref. \onlinecite{wakimoto}). They are
fully consistent with prior observations in Y123
\cite{pailhes1,pailhes3,pailhes2} and unequivocally demonstrate the
universal nature of the bilayer spin excitations, independent of
materials-specific aspects of the crystal structure. They also
greatly extend our capability to correlate the results of INS and
ARPES measurements in order to develop a quantitative description
of the interaction between spin and charge excitations in the
cuprates.

The INS measurements were performed, using the triple axis IN8 at the Institut Laue Langevin,
Grenoble (France), on two $\rm Bi_2 Sr_2 Ca Cu_2
O_{8+\delta}$ single crystal specimens grown by the travelling solvent-floating 
zone method. The first one (OD87) was a monolithic, nearly optimally doped crystal
of mass $\sim$1.5~g and superconducting transition temperature $T_C = 87$
K. Though this sample is lightly overdoped, for convenience, we
refer to it hereafter as optimally doped. The second sample (OD70)
was an array of smaller crystals co-aligned using x-ray Laue
diffraction on aluminium plates. Prior to alignment, the crystals
were annealed at 420 $^\circ$C
under 7 bar of oxygen pressure for 5 days in order
to increase the doping level well into the overdoped regime, so that
$T_C$ was reduced to 70 K. The quality of each crystal was assessed
by inspecting the Laue diffraction pattern and the width of the
superconducting transition (about 3 K)
%CORRECT?
as determined by susceptibility measurements. The mosaic spread of the
array was $\sim 1.6^\circ$, and its total mass was $\sim$3~g.
The INS experimental setup consisted of a pyrolytic graphite (PG)
monochromator and a PG analyser, both set for the (002) reflection.
No collimators were used in order to maximize the neutron flux. The
samples were mounted in a He flow cryostat with the (H,H,0) and
(0,0,L) crystal axes in the scattering plane.
%Tilting of the cryostat allowed the signal at Q=[1.5,1.5,L]
%to be investigated.
%THAT IS IN THE SCATTERING PLANE YOU INDICATE
The wave vector transfer $\bf{Q}$ = (H,K,L) is given in units of
reciprocal lattice vectors $a^* \sim b^* = 1.64 {\rm \AA}^{-1}$
and $c^* = 0.20   {\rm \AA}^{-1}$. In most measurements the final
neutron wave vector was fixed to $k_f =4.1$~\AA$^{-1}$, and a
PG-filter was inserted between the sample and the analyser to cut
higher order contaminations. The energy resolution in this
configuration is $\sim 5$ meV at the energy transfers studied. In
order to extend the energy range, a configuration with  
$k_f=5.5$~\AA$^{-1}$ (with no PG-filter and energy resolution $\sim 8$ meV)
%THAT MUST BE THE RESOLUTION AT THE ELASTIC POSITION, WHICH IS NOT
%RELEVANT HERE: WHAT IS THE INELASTIC RESOLUTION?
was also adopted, and intensity corrections were applied
accordingly.

The dynamical spin susceptibility $\chi(Q,\omega)$ of a bilayer
system can be written as\cite{pailhes1}

\begin{displaymath}
\chi(Q,\omega) = sin^2(\pi zL) \chi_o(Q,\omega) + cos^2(\pi zL)
\chi_e(Q,\omega)
\end{displaymath}

\noindent
where $\chi_o$ and $\chi_e$ denote components that are odd and
even, respectively, under exchange of the layers. $z$ denotes the
layer separation expressed as a fraction of the unit cell
dimension; $z=0.108$ for Bi2212. The component of the momentum
transfer perpendicular to the layers, $L$, can hence be used to
select either odd or even components of $\chi$.

\begin{figure}[htbp]
\begin{center}
\includegraphics[width=12cm,angle=0]{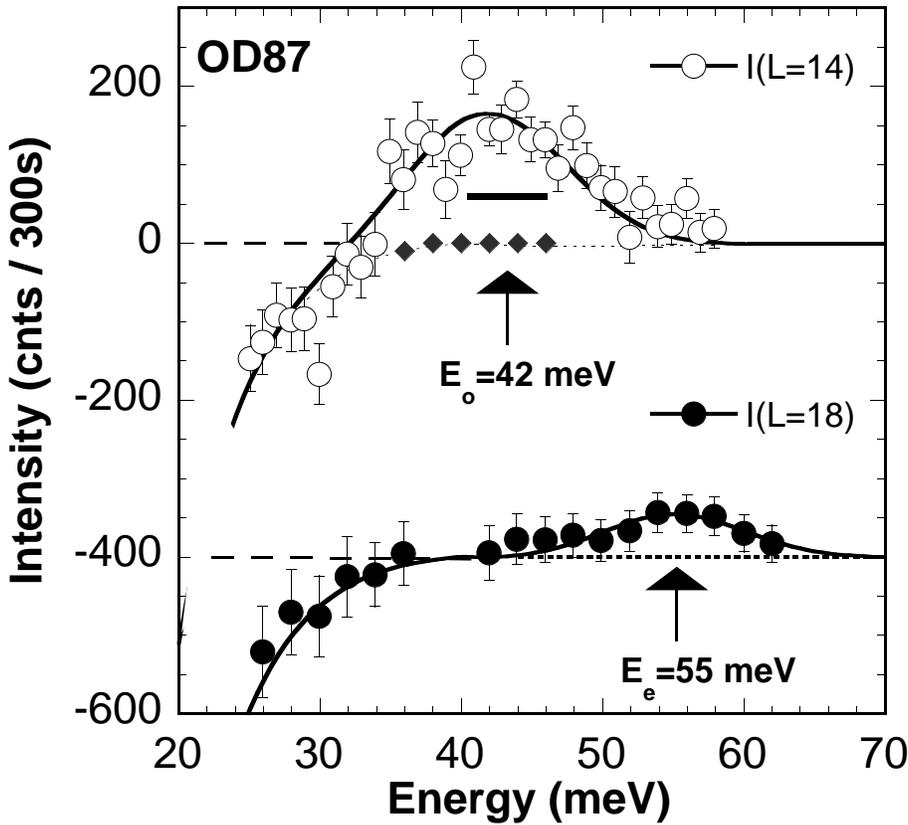}
\caption{Resonant magnetic modes in nearly optimally doped $\rm
Bi_2 Sr_2 Ca Cu_2 O_{8+\delta}$ ($T_C$=87 K) (OD87) measured at ${\bf Q}
=(0.5,0.5,L)$ in the (a) odd channel ($L$=14) and (b) even channel
($L$=18). The signal is obtained by subtracting the intensity at
T=5K from the intensity above $T_C$. The full lines are the
result of a fit by a Gaussian profile on top of a background
(dotted lines). The shape of this background is given by the difference 
between the phonon populations at the two temperatures, yielding negative
values at low energy. The diamond symbols correspond
to the background level obtained by constant energy scans. Mn stands for monitor.
} \label{fig1}
\end{center}
\end{figure}

\begin{figure}[htbp]
\begin{center}
\includegraphics[width=12cm,angle=0]{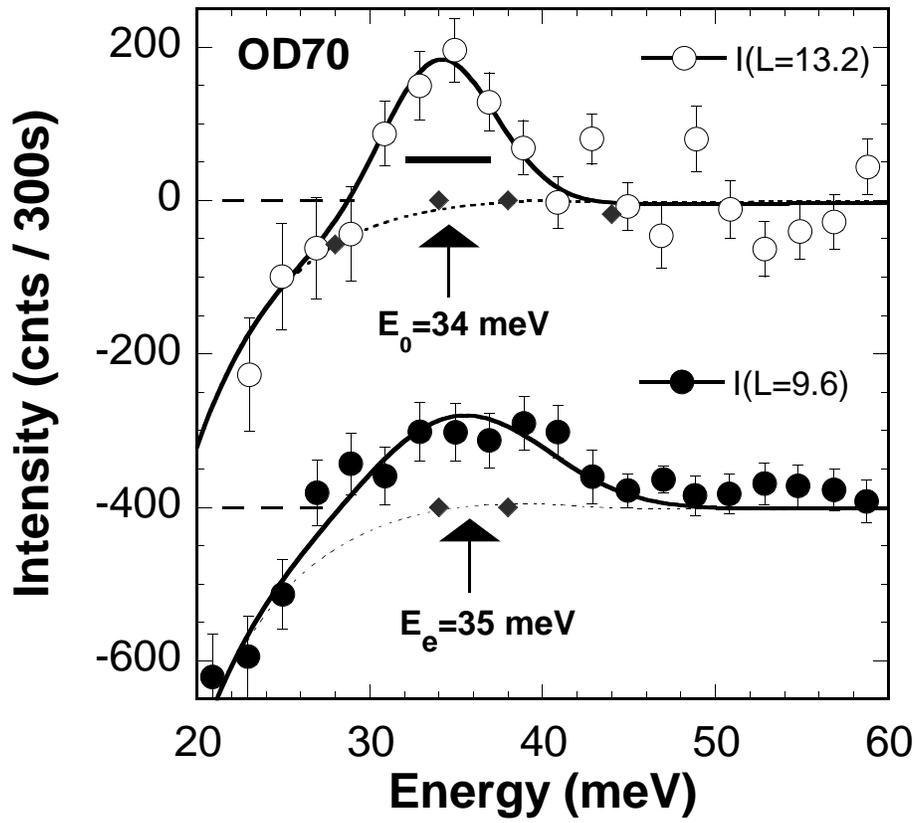}
\caption{Same as Fig. \ref{fig1} in  OD70, highly overdoped $\rm Bi_2 Sr_2
Ca Cu_2 O_{8+\delta}$ ($T_C$=70 K), in (a) odd channel ($L$=13.2) 
and (b) even channel ($L$=9.6). Note that the magnitude of the odd mode
in absolute units is actually $\sim$~3 times weaker in OD70 than in OD87. 
An accurate value cannot be given due to uncertainties in the calibration
procedure. }
 \label{fig2}
\end{center}
\end{figure}

In order to extract the magnetic contribution to the neutron
scattering cross section, we followed procedures established in
prior work on Y123 and Bi2212 \cite{ybco,fong,he}. In particular, it
was shown that the magnetic resonant mode is sharply defined only in
the superconducting state. The mode can thus be experimentally
identified as a sharp enhancement in the difference of the neutron scattering 
intensity between a low temperature (5K) and a temperature just above the 
superconducting transition temperature. Figures \ref{fig1} and \ref{fig2}
%ADD LABELS a,b
show const-{\bf Q} scans at the in-plane wave vector transfer ${\bf
Q}_{\|}=(0.5, 0.5)$, where the spectral weight of the magnetic
resonant mode is known to be maximum. The out-of-plane component of
the wave vector transfer was adjusted such that only odd excitations
contribute to the signal. Both data sets indeed reveal the
characteristic low-temperature enhancement of the cross section
associated with the magnetic resonant mode \cite{ybco,fong,he}.
In order to ascertain the magnetic origin of the difference signal, the
scans were repeated at ${\bf Q}_{\|}= (1.5,1.5)$, keeping $L$ fixed.
At this position, a substantial reduction of the signal was
observed. As phononic signals generally increase in higher Brillouin
zones, this indicates that the signal is of magnetic origin. In both
samples, the observed signals also exhibit other characteristic
signatures of the magnetic resonant mode. Notably, the intensity
enhancement is maximum around ${\bf Q}_{\|}=(0.5, 0.5)$ (not shown),
and the intensity is gradually reduced upon heating and vanishes in
an order-parameter-like fashion at the superconducting transition
temperature (Fig. 3).

A comparison between the magnetic resonant modes in the odd
channel of optimally doped and highly overdoped Bi2212 reveals
interesting information. First, the mode energy decreases with
doping. The energy of the mode in OD87, $E_r^o
= 42$ meV, is consistent with prior work on optimally doped Bi2212.
In the highly overdoped sample $E_r^o = 34$ meV, substantially lower
than that determined in a previously studied, slightly overdoped
sample /cite{he}, but consistent with the relationship $E_r^o = 5.4 k_B T_C$
established on the basis of extensive work on Y123 and Bi2212
\cite{ybco}. The new data therefore underscore our conclusion that
this relation holds generally in the cuprates, and extend its range
of validity into the highly overdoped regime.

Another interesting observation concerns the energy width of the
resonant mode. In the optimally doped sample, the energy width of
the resonant mode is much broader than the instrumental resolution.
This agrees with prior work on optimally doped and slightly
overdoped samples \cite{fong,he}. Due to the high quality of our
samples, combined with the fact that samples from different origins
exhibit the same broadening, $\Delta_{\omega}\sim$11~meV after a deconvolution of 
the energy resolution, it is unlikely that this broadening is of 
extraneous origin (arising, for instance, from a macroscopic
oxygen concentration gradient). Remarkably, the odd peak in
the overdoped sample is sharper, yielding  $\Delta_{\omega}\sim$6~meV 
after the resolution deconvolution. This observation mirrors the doping 
dependence of the inhomogeneity of the superconducting energy gap extracted from
scanning tunnelling spectroscopy data on Bi2212 \cite{mcelroy}. This
inhomogeneity arises from nanometer-sized patches with different SC gap
magnitudes, which follow the distribution of oxygen dopant ions. In
the overdoped range, the distribution of gap amplitudes is significantly
reduced\cite{jlee}. The parallel evolution of the width of (bulk-sensitive)
neutron data on the magnetic resonant mode suggests that the SC gap
distribution is not a pure surface phenomenon. In Y123, however,
the resonant mode is resolution-limited even at optimum doping
\cite{ybco}. The inhomogeneity is thus not generic to the cuprates,
and it is not a precondition for high-temperature superconductivity.

We now turn to the measurements in the even channel, which is probed
for $L=n/z$ with $n$ integer (Figs. 1b and 2b). At optimum doping,
the signal at this position (determined in the same way as the odd
signal discussed above) exhibits an enhancement in the
superconducting state centered at an energy transfer of $E_r^e =54$
meV. Within the experimental error, this is identical to the energy
of the even resonant mode recently identified in optimally doped
Y123 \cite{pailhes3}. In the highly overdoped sample, a similar
superconductivity-induced enhancement is observed around $E_r^e=35$ meV. 
As in Y123, the even mode exhibits the same characteristics
as the odd mode, including the temperature dependence of the
intensity with its onset at $T_C$ (Fig. 3). The even-odd splitting
is thus dramatically reduced in the highly overdoped regime. This
confirms and extends the trend established on the basis of data on
overdoped Y123 \cite{pailhes1,pailhes2}. In the overdoped sample,
the even mode exhibits a larger energy linewidth compared with its 
odd counterpart. This feature is actually also observed in Y123 at all
doping\cite{pailhes1,pailhes3,pailhes2}. The similarity of
data on two distinct families of cuprates allows us to rule out
materials-specific disorder as the origin of this broadening, so
that an intrinsic mechanism appears to be at work. 

%One possibility
%is that the high-energy tail of the mode originates from the edge of
%the particle-hole continuum, whose spectral weight is also expected
%to be enhanced in the superconducting state. The even mode would
%then appear broader because it is closer to the continuum, so that
%the collective mode and the edge of the continuum cannot be
%resolved. However, further work is required to validate this scenario.

As also observed in  Y123\cite{pailhes1,pailhes3,pailhes2}, the intensity 
ratio of both modes intensities changes significantly with doping. 
More quantitatively, we have extracted the energy-integrated spectral weights 
of odd and even modes, $W_r^{o,e}$\cite{pailhes2}. 
Using the fits of Fig. \ref{fig1} and Fig. \ref{fig2} and taking into account 
the fact that the scattering intensity is further weighted by
the magnetic form factor (which is $L$-dependent\cite{pailhes1}), 
one deduces a ratio $W_r^{o}/W_r^{e}$ varying from $\sim$~2.8 in OD87 
to $\sim$~1.5 in OD70. As discussed in Refs. \onlinecite{pailhes2,millis}, 
an estimate of the threshold for
particle-hole excitations in the superconducting state, $\omega_c$,
can then be obtained in the framework of the spin exciton model, where
the spectral weight is approximately proportional to the binding
energy of the resonant mode: $W_r^{o,e}\propto (\omega_c -
E_r^{o,e})/\omega_c$. Fig. 4 provides a synopsis of the continuum
threshold extracted in this way on the Bi2212 samples investigated
here, along with analogous data on Y123. It is gratifying to see
that the data on both systems are consistent. They are also
consistent with the maximum superconducting energy gap extracted
from ARPES\cite{mesot}
%REFERENCE
and electronic Raman scattering (ERS)\cite{raman}
%REFERENCE
data also shown in Fig. 4. This finding is important, because
neutron scattering is a bulk probe, while ARPES and ERS are
sensitive to surface preparation. It also provides further
reassurance of the validity of the spin exciton model.

In conclusion, our experiments have shown that the magnetic
excitations in Y123 and Bi2212 evolve in a strikingly similar way,
despite the different crystal structures and the presence of the
electronically active CuO chains in Y123. This represents a
significant step in the quest for a generic spin response of the
cuprates. We were also able to establish parallels between the
doping evolution of the width of the magnetic resonant mode and the
gap disorder recently observed by scanning tunnelling spectroscopy on
Bi2212 surfaces, indicating that at least part of this effect may be
also representative of the bulk of Bi2212.

This work was supported in part by the Deutsche
Forschungsgemeinschaft, Grant No. KE923/1–2 in the consortium
FOR538.

%\begin{table}[htdp]
%\begin{center}
%\begin{tabular}{|c|c|c|c|c|c|c|}
%\hline
% sample & T$_c$ (K) & E^o$_r$$(meV) & E^e$_r$$(meV) & \omega$_c$$(meV)& References\\
%\hline
%$ {  \rm  Bi_2 Sr_2 Ca Cu_2 O_{8.21}$ & 70 & 34 & 35 & 39 & here\\
%\hline
% ${  \rm Bi_2 Sr_2 Ca Cu_2 O_{8.19}$  &  83 & 38 & NM & NA & \cite{fong}\\
%\hline
%${  \rm  Bi_2 Sr_2 Ca Cu_2 O_{8.18}$  &  87 & 42 & 55 & 59 & \cite{fauque}\\
%\hline
%${ \rm Bi_2 Sr_2 Ca Cu_2 O_{8.16}$  &  91 & 45 & NM & NA & \cite{he}\\
%\hline
% \end{tabular}
%\caption{\label{table1}List of Bi2212 systems with different doping levels in which
%the magnetic resonant modes have been observed. The doping level and the crital temperatures are stated for each case.
% In slightly overdoped and optimally doped Bi2212 the even channel has not been measured (NM). References
%are given where the samples have been described in previous neutron scattering
%studies. }
%\end{center}
%\end{table}

%Modify labels to $E^{e}$ and $E^{o}$ to be consistent with the text

%Modify labels to $E^{e}$ and $E^{o}$ to be consistent with the text

\begin{figure}[htbp]
\begin{center}
\includegraphics[width=12cm,angle=0]{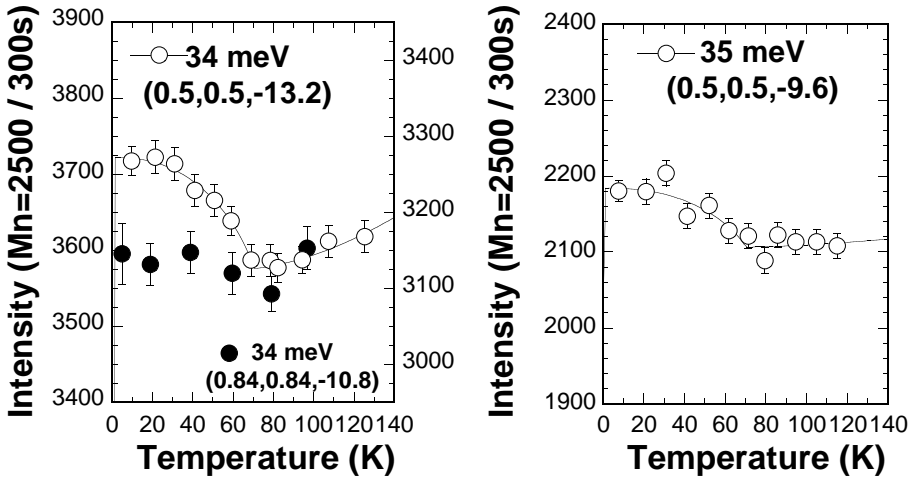}
\caption{Temperature dependence of the peak intensity of both odd
(left panel) and even (right panel)  resonant modes 
(empty symbols) in OD70. The
solid symbols show the temperature evolution of the intensity at a
background point away from the magnetic wave vector. The solid lines
are the results of fits to the data obtained using a power law in
the superconducting state on top of a phononic background.
% governed by the Bose population factor in the normal state.
} \label{fig4}
\end{center}
\end{figure}

\begin{figure}[htbp]
\begin{center}
\includegraphics[width=12cm, angle=0]{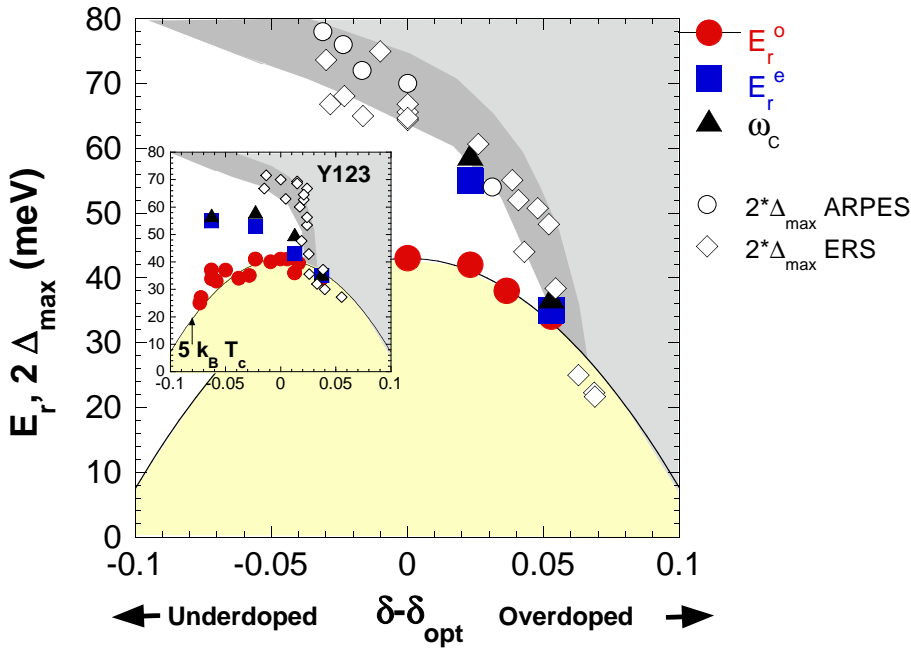}
\caption{Evolution of the magnetic resonant mode energy as a
function of the hole concentration $\delta$ in the Bi2212 system.
(see Ref. \onlinecite{pailhes2} for the definition of $\delta$ ).
The full symbols represent the odd ($E_r^{o}$) and even ($E_r^{e}$)
modes, and the threshold energy ($\omega_c$)as explained in the
legend. The open symbols are the energy gap measured by ARPES\cite{mesot}
and by Electronic Raman Scattering (ERS)\cite{raman}.
%WHERE WERE THE DATA TAKEN FROM? INSERT REFERENCE
The inset shows the equivalent diagram for Y123
\cite{pailhes2}.} \label{fig5mod}
\end{center}
\end{figure}

\end{document}